# Temporal Trends of Intraurban Commuting in Baton Rouge 1990-2010[1]

Yujie Hu[1], Fahui Wang[1]

[1]Department of Geography & Anthropology, Louisiana State University, Baton Rouge, LA 70803, USA

**Abstract**

Based on the 1990-2010 CTPP data in Baton Rouge, this research analyzes the temporal trends of commuting patterns in both time and distance. In comparison to previous work, commuting length is calibrated more accurately by Monte Carlo based simulation of individual journey-to-work trips to mitigate the zonal effect. First, average commute distance kept climbing in 1990-2010 while average commute time increased in 1990-2000 but then slightly dropped toward 2010. Secondly, urban land use remained a good predictor of commuting pattern over time (e.g., explaining up to 90% of mean commute distance and about 30% of mean commute time). Finally, the percentage of excess commuting increased significantly in 1990-2000 and stabilized afterwards.

## Introduction

Commuting is a daily human mobility behavior for employed individuals, and has significant influence on society. Commuting is strongly connected with some practical issues on which many public policies concentrate (Sultana and Weber 2014). It is a major contributor to traffic congestion, air pollution and greenhouse gas emissions. Clearly, the most congested period in a day occurs at the morning/afternoon commuting peak, even though commuting only represents 20-25% share of all-purpose trips in the U.S. (Horner 2004). In line with the worsened traffic congestion, commuters are spending more time in their daily commute trips all across the U.S. For example, one-out-of-twelve U.S. workers in 2001 spend an hour or more in their one-way commute trips, compared to one-out-of-twenty in 1995. It was also reported that the average commute speed declined about 10 percent in midsize metropolitan areas from 1990 to 2009 (http://nhts.ornl.gov/2009/pub/stt.pdf, p.49). Among all economic sectors, transportation (including commuting) is the "second-largest contributor to total U.S. emissions" (next to industry) but with the fastest growth rate (www.state.gov/documents/organization/140636.pdf, p.36). Understanding the *temporal change of commuting* and its underlying causes is a step toward the larger goals of traffic congestion mitigation and carbon emission control.

---

[1] This is a preprint of: Hu, Y., & Wang, F. (2016). Temporal trends of intraurban commuting in Baton Rouge, 1990–2010. *Annals of the American Association of Geographers*, *106*(2), 470-479. https://doi.org/10.1080/00045608.2015.1113117

**Background and Related Literature**

Commuting varies across areas and by socio-demographic groups. For understanding its variability, some focus on spatial factors (Horner 2004), and others emphasize aspatial factors such as race (Kain 2004), wage (Wang 2003) or income (Horner and Schleith 2012), and gender (Kwan and Kotsev 2015). In other words, commuting may be explained by where they are and who they are (Wang 2001). Due to limited space, this paper focuses on the former. The latter will be examined elsewhere.

Specifically, researchers with a spatial perspective have a sustaining interest in explaining intraurban variation of commuting by *land use pattern*. In essence, commuting is for a worker to overcome the spatial barrier from home to workplace, and therefore explanation of commuting naturally begins with a focus on the spatial separation of resident (population) and employment locations. Past attempts include modeling how far a residential location is from a job concentration area such as the CBD and/or subcenters (Wang 2000), or from the overall job market such as job accessibility (Wang 2003), or measuring the need of commuting beyond a local area (e.g., captured by the jobs-housing balance ratio) (e.g., Sultana 2002). However, there is no shortage of doubters on whether commuting could be predicted by urban land use pattern (e.g., Giuliano and Small 1993).

*Excess commuting* is another line of research closely related to the paradigm of interrelatedness between land use and commuting. It is the proportion of actual commute over minimum (optimal or required) commute when assuming that people could freely swap their homes and jobs in a city (Hamilton 1982, White 1988). Instead of focusing on the variation of commuting across areas, it highlights how much overall commuting could be reduced based on the above assumption. In other words, the concept captures the potential (or lack of potential) for a city to optimize commuting without altering the existing land use, and to some extent, reflects efficiency in its land use layout. However, excess commuting was mostly based on a homogeneous work group assumption, and some studies made efforts to relax such assumption by considering individual utility needs, household structure, gender, employment class and other constraints (Cropper and Gordon 1991, Kim 1995, Horner 2002, Horner 2004, Yang 2008). Some studies then extended the idea of measuring minimum commute and proposed other commuting efficiency metrics, such as random commute, maximum commute, and proportionally matched commute (Horner 2002, Yang 2008, Murphy and Killen 2011).

A few recent studies examined the *temporal change of commuting patterns* and related policy implications. Horner (2007) explained the spatial-temporal pattern of intraurban commuting (mileage and multiple commuting efficiency metrics) from the jobs-housing balance perspective in Tallahassee of Florida from 1990 to 2000. Similarly, Chen et al. (2010) investigated the change of commuting patterns from analyzing the residential and employment distributions in Central Texas 1990-2000. Yang (2008) examined the temporal change of excess commuting in Boston and Atlanta 1980-2000, and concluded that urban structure alone could lead to change in excess commuting when controlling individuals' preferences.

Finally, *measures of commute length* merits some discussion. Commuting studies often use commute time as it is directly available from survey data; however, distance could provide a more consistent measure of commuting length (Sultana and Weber 2007). Among the few studies on commute distance, most used a zonal centroid-to-centroid approach to reconstruct either Euclidean or network distance (e.g., Wang 2001, Yang 2008). Such an approach could bias the estimate particularly in large zones and also by omitting intra-zonal distance (Hewko et al. 2002). For this reason, Horner and Schleith (2012) used a small analysis unit such as census block in order to mitigate the scale effect. However, commuting data at the block level is not widely available for most cities in the U.S. Different unit scales and unit zone definitions cause inconsistency in analysis results, particularly common in comparison analysis over time, and thus lead to the *modifiable areal unit problem (MAUP)* (Niedzielski et al. 2013). Some most recent studies have shown great promises of using the GPS data and activity-travel surveys of individual trip makers in commuting studies (Shen et al. 2013, Kwan and Kotsev 2015). However, such data may not be representative of all commuters and are also not universally available. More accessible and accurate measures of commute length remain very much needed.

This paper examines the aforementioned issues that have been central to the study of commuting and urban structure. The contributions of our research can be highlighted in three aspects. Foremost, we measure commute length by Monte Carlo based simulation of individual resident workers and jobs and the trips between them, a significant improvement over the zonal-level centroid-to-centroid approach, to mitigate the aforementioned aggregation error and scale effect. Secondly, the new and longer temporal trends for commuting are detected by analyzing the newly-available 2006-2010 Census Transportation Planning Package (CTPP) data. Finally, when defining excess commute, the optimal commuting pattern is now formulated as an integer programming problem based on simulated individual trips instead of the conventional linear programming approach for zonal-level trips.

**Data Sources and Commute Distance Calibration**

The study area is East Baton Rouge Parish in Louisiana (hereafter simply referred to as Baton Rouge). With a population of 440,000 in 2010, it is the core of Baton Rouge Metropolitan Area (other surrounding parishes are mostly rural). Parish is a county equivalent unit in Louisiana. The major data sources are: the 1990 and 2000 CTPPs extracted from the long form decennial census (www.transtats.bts.gov), and the most recent 2006-2010 CTPP based on the 5-year American Community Survey (ctpp.transportation.org/Pages/5-Year-Data.aspx). All CTPPs consist of three parts: Part 1 on residential places (e.g., number of resident workers, worker breakdowns by transportation mode or by wage range), Part 2 on workplaces (e.g., number of jobs), and Part 3 on journey-to-work flow (e.g., number of commuters and mean travel time on OD trips). For simplicity, *the 2006-2010 data and corresponding analyses are hereafter referred to as 2010.*

The CTPP data in Baton Rouge used different area units over the years: traffic analysis zone (TAZ) in 1990, multiple zones in 2000 (census tract, census block group, and TAZ for Parts 1 and 2, only census tract for Part 3), and census tract and TAZ in 2010. We chose census tract as the unified unit. Luckily in Baton Rouge, the 1990 TAZs were mostly components of census tracts for easy aggregation with only very few minor adjustments. There were 85, 89 and 91 census tracts in Baton Rouge in 1990, 2000 and 2010, respectively. As shown in Figure 1, the major job concentrations of the study area are (1) downtown (CBD), (2) the largest employer Louisiana State University (about 3 miles south of the CBD), and (3) a large area with several major hospitals and health clinics (3-5 miles southeast of downtown). The pattern is generally consistent with that in 2000 (Antipova and Wang 2010, Antipova et al. 2011).

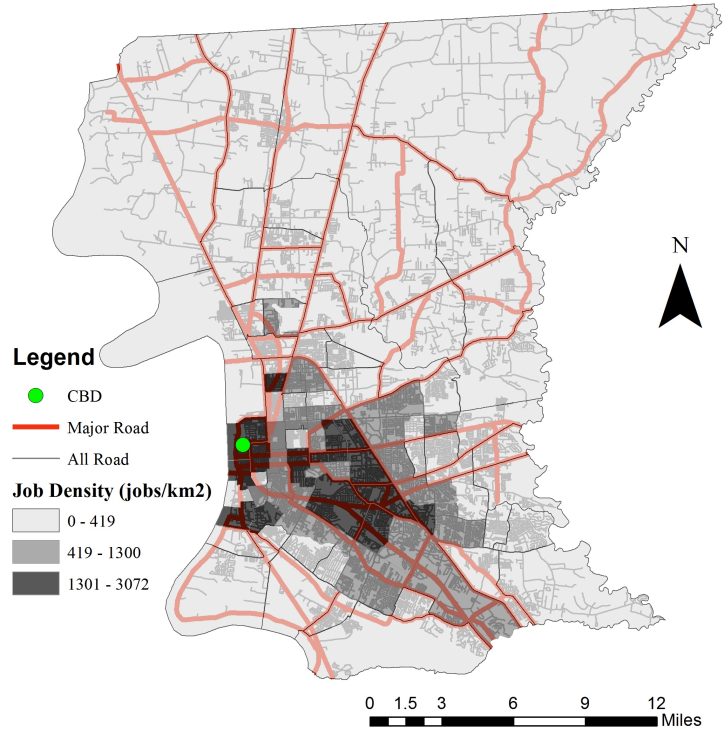

Figure 1. Employment density in Baton Rouge 2010

The CTPP provides information on commute time but not on commute distance. We use *Monte Carlo simulation* to obtain journey-to-work trips between individual points. Briefly, the first task is to randomly generate points of resident workers and jobs within tracts so that their total numbers at the zonal level are proportional to the observed patterns of resident workers and jobs in the CTPP. The second task is to pair the origins (workers) and destinations (jobs) to form OD trips that follow a discrete frequency distribution that is also consistent with the reported journey-to-work flows in the CTPP, and then measure the network distance for each OD trip. For more technical detail, see Hu and Wang (2015).

By doing so, the zonal journey-to-work flow is disaggregated into individual trips. It permits more accurate estimation of commute distances and mitigates the aggregation error and zonal effect. For example, the average commute distance in Baton Rouge in 2000 was 5.95 miles by the zonal centroid-to-centroid approach, and 6.17 miles by our simulation-based approach. The longer estimates by the simulation technique are validated through a one-tailed t test. In addition, the centroid-to-centroid approach may assign the intrazonal travel distance uniformly as 0 whereas our approach yields a more realistic measure that varies with the zone size (e.g., 7.94 miles in the largest tract in the northeast corner). This is especially important in excess commuting analysis because the "optimal" commuting pattern based on the zonal data often includes a large number of intrazonal trips when assuming a zero intrazonal commute. Our approach yields the optimal pattern with more inter-zonal trips when a trip is shorter for a resident worker travels to a job in a neighboring tract than within the same tract, which is more realistic.

**Overall Temporal Trend of Commuting**

People commute by multiple modes. As shown in Table 1, the majority commuted by auto (including drove-alone and carpool), and the percentage was steady over time (i.e., 94-95%). Therefore, we estimated commute distance based on the road network without taking transit, bicycle or pedestrian routes into consideration. Mean commute distance (time) in a tract is the average of travel distances (times) from this tract to all tracts weighted by corresponding number of commuters. Specifically, a tract's mean commute time was based on the zonal commuter flow matrix and corresponding travel time reported in the CTPP. However, mean commute distance, while not reported, was based on the Monte Carlo simulation of individual trips.

Table 1. Modal splits and commute lengths in Baton Rouge 1990-2010

| Year | Modal splits | | | | Mean commute distance (mile) | Mean commute time (min) |
|---|---|---|---|---|---|---|
| | Drove-alone | Carpool | Public transit | Others[1] | | |
| 1990 | 82.35 | 11.76 | 1.29 | 4.60 | 5.95 | 16.73 |
| 2000 | 83.16 | 11.91 | 1.40 | 3.54 | 6.17 | 18.73 |
| 2010[2] | 83.77 | 11.08 | 1.75 | 3.40 | 6.25 | 17.98 |

[1] Others include taxi, motorcycle, bicycle, walk, etc.

[2] Based on 2006-2010 CTPP, 2010 is used here and all other tables for simplicity.

Also as shown in Table 1, the mean commute distance on average increased steadily from 5.95 mile in 1990 to 6.17 mile in 2000 and further to 6.25 mile in 2010. This is consistent with some previous studies (Levinson and Kumar 1994, Cervero and Wu 1998), and reflects a long standing trend of more workers moving farther from their jobs either to search farther for jobs for maximizing their earnings or to move their residences for better housing. However, the increasing rate was higher in 1990-2000 than in 2000-2010. One possible reason might be the economic recession that began in 2008 (Horner and Schleith 2012).

The mean commute time for overall population increased from 1990 to 2000 and then declined to 2010. Other studies also found that the average commute time stayed stable or even dropped over time (e.g., Kim 2008). They ascribed the declining time to *co-location of jobs and housing*, i.e., people relocate their residence or jobs to cut back commute time as traffic becomes more congested. An increasing use of suburban roads that are usually newer and wider than roads in central city makes it achievable to commute longer distance in a shorter duration. Taking drove-alone for example, the implied average commuting speed was 24.3 mph in 1990, dropped to 21.9 in 2000 and only recovered slightly to 22.8. Therefore, a small increment in commute distance from 1990 to 2000 came with a much larger climb in commute time, and it was only after 2000 that the co-location theory became relevant and led to a small drop in commute time. The significant climb of commute time for 1990-2000 might reflect people's increasing endurance of long commutes or traffic congestion that grew much more rapidly than people could adapt (Levinson and Wu 2005), and the re-location adjustment came afterwards.

**Commuting and Land Use Patterns**

There has been a long tradition of attempts to explain intraurban variability of commuting by land use patterns. The analysis begins by examining the impact of the CBD with the highest employment concentration. Table 2 shows that *distance from the CBD* (denoted by $D_{CBD}$) explained the variation of mean commute distance across census tracts by 78% in 1990, 68% in 2000 and 63% in 2010. The declining explaining power was attributable to dispersion of jobs beyond the CBD area (e.g., percentage of workers commuted to the area within a 3-mile radius of the CBD was 42% in 1990, 32% in 2000 and 31% in 2010). The effect of distance from CBD remained significant on mean commute time in all models (Table 3), but much weaker than on mean commute distance (Table 2). The lower performance of regression models on commute time was attributable to the non-uniform modal distribution across tracts. In fact, for drove-alone commuters alone, the pattern of mean commute time was largely consistent with that of mean commute distance, and the corresponding regression models yielded $R^2$ = 0.60, 0.41 and 0.50 in 1990, 2000 and 2010, respectively (details not reported here). Adding the square term $D_{CBD}^2$ did not improve the explaining power of the regression models and thus was not included.

Table 2. Regression models of mean commute distance across census tracts 1990-2010

|  | 1990 | 2000 | 2010 | 1990 | 2000 | 2010 | 1990 | 2000 | 2010 |
|---|---|---|---|---|---|---|---|---|---|
| Intercept | 2.40*** | 2.91*** | 3.01*** | 10.37*** | 14.87*** | 13.65*** | 0.88*** | 0.99*** | 1.10*** |
|  | (10.03) | (10.52) | (9.93) | (29.14) | (20.75) | (21.34) | (4.60) | (7.06) | (5.74) |
| $D_{CBD}$ | 0.57*** | 0.50*** | 0.48*** |  |  |  |  |  |  |
|  | (17.38) | (13.67) | (12.22) |  |  |  |  |  |  |
| JWR |  |  |  | -3.88*** | -10.38*** | -8.05*** |  |  |  |

|       |       |       |       | (-14.07) | (-8.35) | (-7.57) |       |       |       |
|-------|-------|-------|-------|----------|---------|---------|-------|-------|-------|
| JWR$^2$ |     |       |       | 0.37$^{***}$ | 2.60$^{***}$ | 1.75$^{***}$ |       |       |       |
|       |       |       |       | (13.93)  | (5.33)  | (4.36)  |       |       |       |
| JobP  |       |       |       |          |         |         | 0.67$^{***}$ | 0.67$^{***}$ | 0.64$^{***}$ |
|       |       |       |       |          |         |         | (29.44) | (40.26) | (29.22) |
| No. obs. | 85 | 89  | 91    | 85       | 89      | 91      | 85    | 89    | 91    |
| R$^2$ | 0.78  | 0.68  | 0.63  | 0.72     | 0.78    | 0.76    | 0.91  | 0.95  | 0.91  |

Note: t-statistics are in parentheses; $^{***}$significant at the 0.001 level.

Secondly, we look into the *jobs-housing balance* hypothesis proposed by Cervero (1989). An imbalanced area has far more resident workers than jobs, and thus more workers need to commute outside of the area for their jobs and tend to incur more commuting. Following Wang (2000), this research used the floating catchment area method to define a circular area around each census tract centroid, and calculated the jobs-to-workers-ratio (JWR) within each catchment area. A higher JWR implies less need of commuting beyond the catchment area and thus is expected to correlate with less commute. We experimented with radii ranging 1.5-7.5 miles and settled with 5 miles for its best explaining power.

As an example, Figure 2 shows the relationship between mean commute distance and time versus JWR in 2010. Both display a quadratic trend, but the trend is much clearer for distance than time. This is confirmed by a regression model with the added square term of JWR, as reported in Tables 2 and 3. In 1990, 2000 and 2010, the negative sign of JWR and the positive sign of JWR$^2$ indicate that mean commute distance or time at the tract level declined with JWR, but the declining slope got flatter in higher-JWR areas. Both terms are statistically significant in all models. The models for mean commute distance in Table 2 performed well with R$^2$ that was 0.72 in 1990, peaked at 0.78 in 2000 and dropped slightly to 0.76 in 2010. The models for mean commute time in Table 3 also confirmed the quadratic trend in all years though with lower R$^2$ (0.21, 0.26 and 0.35 in 1990, 2000 and 2010, respectively). In short, the results confirm the importance of jobs-housing imbalance in affecting commuting pattern, much more significant in Baton Rouge than large metropolitan areas reported in other studies (e.g., Wang 2000).

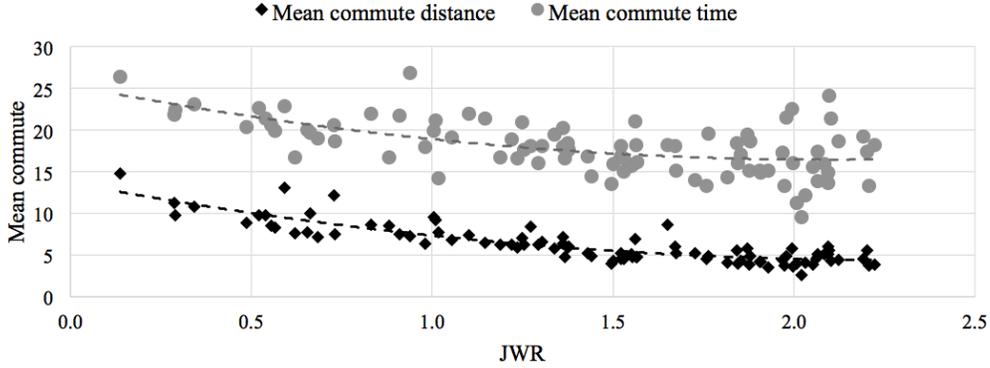

Figure 2. Mean commute distance and time vs. jobs-to-workers ratio (JWR) in 2010

Table 3. Regression models of mean commute time across census tracts 1990-2010

|  | 1990 | 2000 | 2010 | 1990 | 2000 | 2010 | 1990 | 2000 | 2010 |
|---|---|---|---|---|---|---|---|---|---|
| Intercept | 14.72*** (26.82) | 16.96*** (23.06) | 15.29*** (25.01) | 19.69*** (28.25) | 27.86*** (13.51) | 25.36*** (17.12) | 13.32*** (20.44) | 14.42*** (16.12) | 13.63*** (21.15) |
| $D_{CBD}$ | 0.32*** (4.28) | 0.27** (2.80) | 0.40*** (5.04) |  |  |  |  |  |  |
| JWR |  |  |  | -2.51*** (-4.65) | -12.26*** (-3.43) | -8.45*** (-3.43) |  |  |  |
| $JWR^2$ |  |  |  | 0.21*** (4.12) | 3.60* (2.57) | 1.99* (2.15) |  |  |  |
| JobP |  |  |  |  |  |  | 0.25*** (5.74) | 0.30*** (5.22) | 0.31*** (7.41) |
| No. obs. | 85 | 89 | 91 | 85 | 89 | 91 | 85 | 89 | 91 |
| $R^2$ | 0.18 | 0.08 | 0.22 | 0.21 | 0.26 | 0.35 | 0.28 | 0.24 | 0.38 |

Note: t-statistics are in parentheses; *significant at the 0.05 level, **significant at the 0.01 level, ***significant at the 0.001 level.

Either the emphasis on the role of CBD or the jobs-housing balance approach does not consider all job locations in explaining commuting patterns. The *job proximity index* (*JobP*) captures the spatial separation between a worker's residence and all potential job sites (Wang 2003), formulated such as:

$$JobP_i = \sum_{j=1}^{n}(P_{ij}d_{ij}), \qquad (1)$$

$$\text{where } P_{ij} = (J_j d_{ij}^{-\beta}) / \sum_{k=1}^{n} (J_k d_{ik}^{-\beta}).$$

Similar to the notion of Huff (1963) model, the probability of workers residing in zone $i$ and going to work in zone $j$ (denoted by $P_{ij}$) is predicted as the gravity kernel of job site $j$ out of all job sites $k$ (= 1, 2, ..., $n$). Each gravity kernel is positively related to the number of jobs there $J_j$ (or $J_k$) and negatively to the distance or time between them $d_{ij}$ (or $d_{ik}$) powered to the distance friction coefficient $\beta$. Then JobP at zone $i$ is simply the aggregation of all distances (time) $d_{ij}$ with corresponding probabilities $P_{ij}$ over all job sites ($j$ = 1, 2, ..., $n$).

Calibration of JobP requires defining the value for the distance friction coefficient $\beta$ in Equation (1). In this research, the $\beta$ value was computed from the log-transformed regression based on the classic gravity model such as

$$C_{ij} = aW_i J_j d_{ij}^{-\beta}, \tag{2}$$

where $C_{ij}$ is the number of commuters from a tract with $W_i$ resident workers and to a tract with $J_j$ jobs for a distance (time) of $d_{ij}$ (Wang 2015, pp.33). Based on the CTPP data, the derived $\beta$ value was 0.404 in 1990, 0.547 in 2000, and 0.475 in 2010 if the journey-to-work trips were measured in distance; and the $\beta$ value was 0.295 for 1990, 0.353 for 2000, and 0.385 for 2010 if measured in time.

The regression results for mean commute distance and time by JobP are again reported in Tables 2 and 3, respectively. The mean commute distance at the tract level was well explained by JobP with $R^2$ = 0.91, 0.95 and 0.91 in 1990, 2000 and 2010, respectively. This is a significant improvement over other factors, i.e., $D_{CBD}$ and JWR. Similarly, regression models on mean commute time returned lower $R^2$ ranging 0.24-0.38. We also did not add the square term JobP$^2$ as it did not improve the explaining power of the regression models. Note that we also run a series of regression models on mean commuting distance (centroid-to-centroid), and results indicate weaker $R^2$ than the simulation-calibrated distance (due to space limit, results are not shown here). This again demonstrates the value of our simulation approach.

**Temporal Change of Excess Commuting**

Although the land use pattern in Baton Rouge explained the commuting variations to some extent, there was still a proportion of variations unexplained (particularly commute time). Part of the gap could be attributable to the "*excess commuting*" behavior that individuals do not necessarily optimize their journey-to-work trips as suggested by the spatial arrangements of land uses, i.e., jobs and houses. Even if individual workers make every effort to minimize their commuting, the outcome may still differ from what is required by minimizing total commute collectively for the whole study area.

Most studies adopt the Linear Programming (LP) approach by White (1988) to define the minimum commute based on zonal data such as:

$$\text{Min } T_{\min} = \sum_{i=1}^{m}\sum_{j=1}^{n}(x_{ij}d_{ij}) \qquad (3)$$

Subject to:

$$\sum_{j=1}^{n} x_{ij} = W_i \qquad (4)$$

$$\sum_{i=1}^{m} x_{ij} = J_j \qquad (5)$$

where $x_{ij}$ is the (non-negative) optimal number of commuters from a resident worker tract $i$ to a job tract $j$ to be solved, and $d_{ij}$, $W_i$ and $J_j$ are the same as defined in Equation (2).

There are a couple of concerns on the above zonal model. The first is lack of spatial accuracy when using average commute time (distance) between zones. Another involves inconsistent and unreliable strategies for defining intrazonal commute lengths (e.g., some use 0, and others use reported time or estimated distance). Both are especially problematic in large zones, or termed the MAUP as discussed in Horner and Murray (2002). As explained previously, this research uses the Monte Carlo approach to simulate the individual locations for workers and jobs to improve the accuracy of estimation. Index the locations for simulated workers and jobs as $k$ and $l$, respectively. The total number of simulated workers is the same as that of simulated jobs, denoted by $n$. It is formulated as an integer programming problem such as:

$$\text{Min } T_{\min} = \sum_{k=1}^{n}\sum_{l=1}^{n}(x_{kl}d_{kl}) \qquad (1)$$

Subject to:

$$\sum_{l=1}^{n} x_{kl} = 1 \qquad (7)$$

$$\sum_{k=1}^{n} x_{kl} = 1 \qquad (8)$$

Note that $x_{kl}$ is a binary integer variable (=1 when a journey-to-work flow is chosen by the optimization algorithm between a simulated worker location and a simulated job location, and = 0 otherwise). The objective function (6) is to minimize the total commute by $n$ simulated commuters, and Equations (7) and (8) ensure that each worker can be assigned to one unique job and vice versa.

Figure 3 summarizes the results on excess commuting in Baton Rouge 1990-

2010. Once again, the average commute distance in the study area increased steadily from 1990 to 2000 and again to 2010; and the average commute time increased from 1990 to 2000 but dropped slightly to 2010. The portion of excess commuting measured in time was consistently higher than that in distance in corresponding years (67.66% vs. 54.27% in 1990, 75.41% vs. 63.99% in 2000, and 75.34% vs. 63.48% in 2010). This discrepancy is understandable since actual commute time included those by slower transportation modes and the optimal time was estimated solely by drove alone. As the concept of excess commuting is proposed mainly to assess the potential of commuting reduction given the land use pattern of a city, our discussion here focused on the results in terms of commute distance.

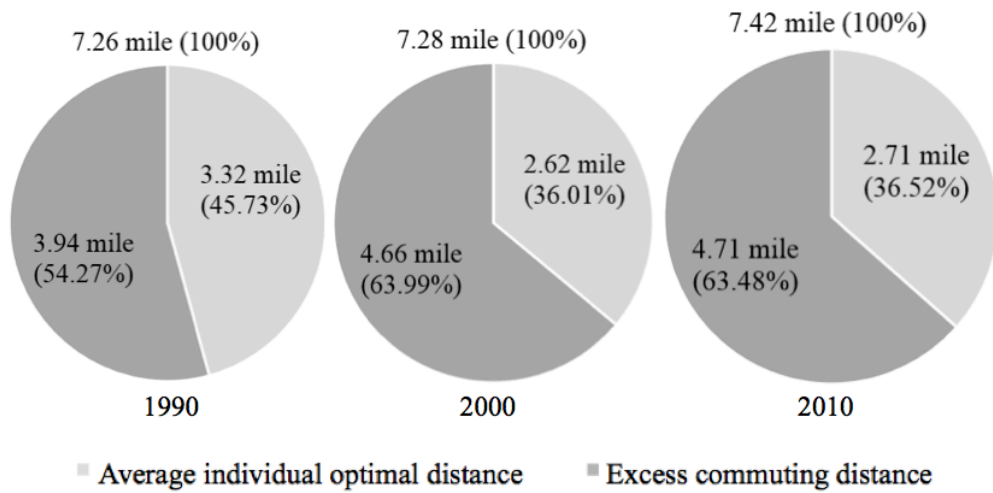

(a)

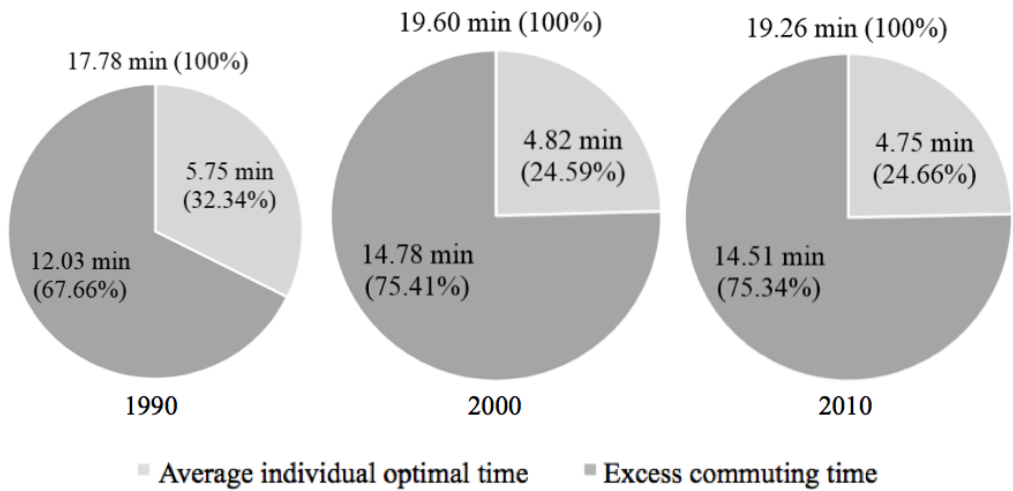

(b)

Figure 3. Excess commuting in 1990-2010: (a) distance, (b) time

    The minimum (required) commute was 3.32 miles in 1990, and dropped to 2.62 in 2000 and inched up slightly to 2.71 in 2010. It suggested that land uses in Baton Rouge might have changed in a way toward more efficiency in terms of commuting need from 1990 to 2000 (e.g., improved proximity between jobs and resident workers in general), and stayed largely stable till 2010. However, the resident workers did not take advantage of the change, and actually increased their trip lengths on average from 7.26 in 1990 to 7.28 in 2000 and then again to 7.42 miles in 2010. This led to the rise of excess commuting distance from 54.27% in 1990 to 63.99% in 2000, and stayed at 63.48% in 2010. Many factors may have contributed to this trend of largely increasing excess commuting, such as an increasing female labor participation rate (and thus more multi-worker households) and a small increase in carpool modal share from 1990 to 2000 (Table 1). A more in-depth discussion on the impact of other aspatial factors is beyond the scope of this paper.

**Conclusion**

This research analyzes the temporal trends of commuting patterns in both time and distance, explains the observed commuting patterns from a spatial perspective (i.e., land use) and measures the extent of excess commuting. Findings are summarized as follows.

    First, mean commute distance steadily increased over time in Baton Rouge, and mean commute time rose along with mean commute distance during 1990-2000, but dropped slightly afterwards. The gap between the two in 2000-2010 was

attributable to a non-uniform modal distribution across tracts. A future research may expand the study area by incorporating neighboring parishes to see whether there are more pronounced changes in the metropolitan area, and use high temporal resolution data to examine the impact of major external factors (e.g., significant fluctuation in gas price) on commuting.

Secondly, on the interrelatedness between intraurban commuting and land use patterns, our research indicates that the spatial variability of mean commute distance in this midsize city can be well explained by distance from the CBD, jobs-housing balance ratio, and even more than 90% by the job proximity index. The models on mean commute time also show improvement over existing studies. The better results may be attributable to improved measure of commute distance and/or a moderate city size in this study. The finding lands support to the effectiveness of planning policies that are aimed at trip reduction by improving jobs–housing balance and job proximity.

Finally, our study finds that Baton Rouge in entirety experienced an increase of excess commuting from 1990 to 2000 in both commute distance and time, and stayed at about the same levels toward 2010. This indicates that the land-use configuration changed in a way that jobs collectively moved closer to residences and thus became better balanced from 1990 to 2000, but the resident workers did not take advantage of that and incurred more excess commuting. The trend of rising excess commuting was largely halted in 2000-2010. The economic downturn beginning in 2008 might be one reason underlying the new trend (Horner and Schleith 2012). The low temporal resolution data (i.e., five-year pooled 2006-2010 CTPP) prevents us from validating this speculation.

On the methodological front, this research proposes a Monte Carlo simulation-based approach for improved modeling of commuting patterns. Granted, a more disaggregated data (e.g., block level) would incur minor aggregation error. However, such detailed data in terms of both spatial and temporal resolution are not always available in commuting studies. Given the more widely available zonal-level commuting data (e.g., CTPP in the U.S.), our approach mitigates the zonal effect and permits more accurate estimate of commute length and subsequently more reliable measure of excess commuting. The formulation of integer programming also has a good potential for wider adoption in modeling the optimal commuting pattern of individual trip makers.